\documentclass[sigconf]{acmart}
\AtBeginDocument{%
  }

\setcopyright{acmlicensed}
\copyrightyear{2026}
\acmYear{2026}
\acmDOI{XXXXXXX.XXXXXXX}
\acmConference[MLCAD '26]{MLCAD '26: ACM/IEEE International Symposium on Machine Learning for CAD}{Sep. 07--09, 2026}{Jeju, South Korea}

\usepackage[dvipsnames]{xcolor}

\usepackage{tabularray}
\usepackage{pifont}
\usepackage{pgf-pie} 
\usepackage{pgfplots}
\pgfplotsset{compat=1.18} %
\usepackage{etoolbox} 
\usepackage{float}
\usepackage{rotating}
\usepackage{tikz}
\usepackage{stfloats}
\usepackage{subfigure}
\usepackage{enumitem}

\usepgfplotslibrary{colorbrewer}

\begin{document}

\title{PCB-QA: Evaluating LLMs over the First Printed Circuit Board Design Question-Answer Dataset}

\author{Sahana Srinivasan}
\orcid{0009-0008-1304-7824}
\affiliation{%
  \institution{UNSW Sydney}
  \country{Australia}
}
\email{sahana.srinivasan@unsw.edu.au}

\author{Benjamin Tan}
\affiliation{%
  \institution{University of Calgary}
  \country{Canada}}
\email{benjamin.tan1@ucalgary.ca}

\author{Benjamin Turnbull}
\affiliation{%
  \institution{UNSW Canberra}
  \country{Australia}}
\email{benjamin.turnbull@unsw.edu.au}

\author{Hammond Pearce}
\affiliation{%
  \institution{UNSW Sydney}
  \country{Australia}}
\email{hammond.pearce@unsw.edu.au}

\begin{abstract}
Large Language Models (LLMs) have demonstrated capabilities in electronic design automation (EDA) for integrated circuits. However, their applications in printed circuit board (PCB) design and analysis tasks remain underexplored. In part, this is due to a (1) a lack of text-based PCB datasets to evaluate LLMs and (2) a lack of methodologies for prompting LLMs with different types of PCB design files. To address this gap, our paper proposes PCB-QA: a manually created questionnaire dataset amounting to 480 question-answer pairs for PCBs, derived from 8 different open-source hardware projects of varying complexities. We examine multiple aspects of PCB designs and cover questions about component connections, datasheet examination, and simulation data obtainable via SPICE. Using our dataset as a benchmark, we prompt LLMs with different representations (and combinations) of PCB design files and record observations. This allows us to measure, for the first time, if LLMs can understand schematics and netlists in their ``native forms" (i.e. graphical PDFs, KiCAD-format design files) or if textual formats are preferred. Including both commercial and open-weight models, we benchmark 4 state-of-the-art LLMs on our dataset, finding that Gemini 3 Flash Preview can answer questions with an accuracy of 93\% using a proposed JSON-based textual format. This demonstrates that text-based PCB design formats can be evaluated by LLMs. Our open-source questionnaire is the first step towards enabling LLM integrations within the PCB design life cycle.
\end{abstract}




\received{29 May 2026}

\maketitle

\section{Introduction} \label{sec:Intro}

\begin{figure}[t]
    \centering
    \includegraphics[width=0.99\linewidth]{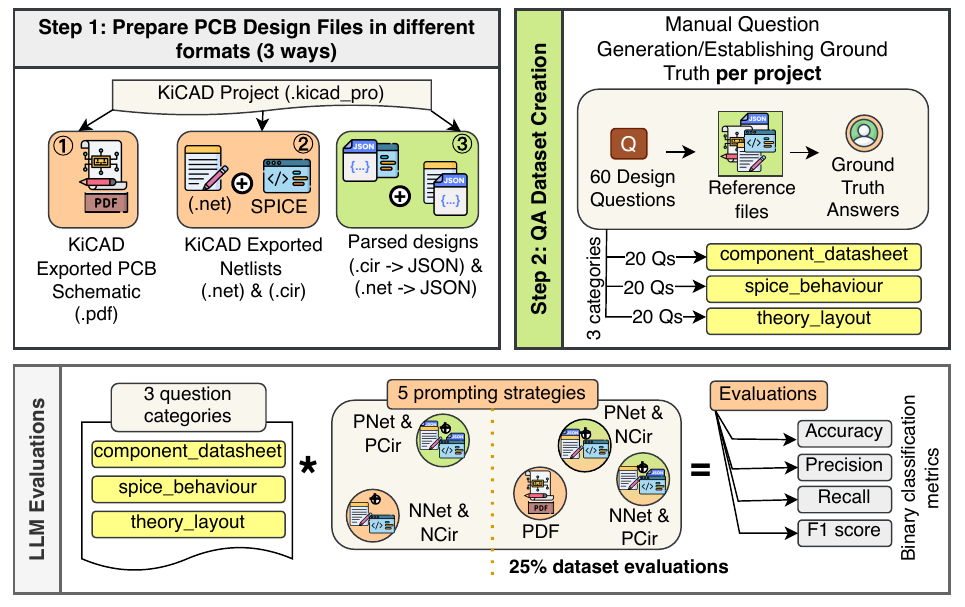}
    \caption{Overview of the methods to prepare the different PCB design format sets \ding{192}, \ding{193} and \ding{194} (Step 1) and to create the question-answer pairs in all 3 categories for each project (Step 2). The dataset is evaluated against some selected LLMs to measure interpretation capabilities using the format sets.}
    \label{fig:main_figure}
\end{figure}

Electronic Design Automation (EDA) has witnessed the integration of Large Language Models (LLMs) for hardware design tasks \cite{vungarala_limca_2025, vungarala_prompt_2025, li_what_2025, lin_pegpt_2025} and embedded software applications, including Verilog generation \cite{liu_rtlcoder_2025, thakur_autochip_2024, wong_vgv_2024, chang_chipgpt_2023} and runtime debugging \cite{li_edadebugger_2025}. 
However, the application of LLMs to printed circuit boards (PCBs) is a field that remains underexplored.
Like integrated circuit internals, which can be modelled in languages such as VHDL and Verilog, PCBs consist of interdependent components and connections.
However, the primary interchange of PCB designs is graphical. %
For example, circuit schematics depict component connections, netlists provide component metadata and nets as graphs, and layout files showcase board geometries. 
These representations do not align naturally with the text-centric architectures of standard LLMs, meaning LLMs may struggle to isolate segments for design and analysis without specialised fine-tuning methods, established sets of rules or tailored prompting strategies \cite{qiu_hwebench_2026}.

At the schematic design stage, current LLM-applied research for PCBs has therefore focused only on interpreting textual design specifications \cite{jansen_words_2023} or generating code-based analog designs \cite{lai_analogcoder_2025, liu_physicsinformed_2024}.
Further works which might use LLMs for other PCB design tasks remain underexplored, with only anecdotal experiences documented.

\begin{table*}[b]
    \centering
    \caption{Summary of related works in the field of PCB datasets and benchmarks. While most existing PCB datasets comprise of images for anomaly detection, our dataset is text-based and can be used by LLMs especially in the early PCB design stages.}
    \resizebox{0.98\linewidth}{!}{
    \definecolor{Concrete}{rgb}{0.949,0.949,0.949}
\definecolor{YellowGreen}{rgb}{0.803,0.921,0.545}
\begin{tblr}{
  row{even} = {Concrete},
  row{1-2} = {c},
  cell{1}{1} = {r=2}{c},
  cell{1}{2} = {r=2}{c},
  cell{1}{3} = {r=2}{},
  cell{1}{4} = {r=2}{},
  cell{1}{5} = {c=3}{c},
  cell{3-11}{3-7} = {c},
  vline{2-7} = {-}{},
  hline{1,3,9} = {-}{0.12em},
  hline{2} = {5-7}{},
}
Work & {Contribution\\Summary} & {Text-\\based?} & {Focused on\\LLM appli-\\-cation?} & {Applied to which stage of PCB manufacturing?} &  & \\
 &  &  &  & {Circuit\\Design} & {Layout and \\Routing} & {Assembly and\\Installation}\\
DeepPCB \cite{tang_online_2019} & {Aligned image pairs with bounding box labels spanning 6 PCB defect categories} & \textcolor{BrickRed}{\ding{55}} & \textcolor{BrickRed}{\ding{55}} & \textcolor{BrickRed}{\ding{55}} & \textcolor{BrickRed}{\ding{55}} & \textcolor{ForestGreen}{\ding{51}}\\
FICS PCB X‑Ray \cite{mehta_fics_2022} & Capture defects beneath solder masks & \textcolor{BrickRed}{\ding{55}} & \textcolor{BrickRed}{\ding{55}} & \textcolor{BrickRed}{\ding{55}} & \textcolor{BrickRed}{\ding{55}} & \textcolor{ForestGreen}{\ding{51}}\\
ElectroVizQA \cite{meshram_electrovizqa_2024} & {Multi-choice Q\&A dataset comprising of text and images to measure multimodal LLM\\performances on analog and digital circuit problems}  & \textcolor{BrickRed}{\ding{55}} & \textcolor{ForestGreen}{\ding{51}} & \textcolor{BrickRed}{\ding{55}} & \textcolor{BrickRed}{\ding{55}} & \textcolor{BrickRed}{\ding{55}} \\
CIRCUIT \cite{skelic_circuit_2025} & {Q\&A dataset to evaluate LLMs on analog circuit knowledge, using diagrams and netlists as inputs} & \textcolor{BrickRed}{\ding{55}} & \textcolor{ForestGreen}{\ding{51}} & \textcolor{BrickRed}{\ding{55}} & \textcolor{BrickRed}{\ding{55}} & \textcolor{BrickRed}{\ding{55}} \\
PCBBench~\cite{li_pcbbench_2025} & {LLM-driven PCB layout, placement and routing benchmark with 170 included samples} & \textcolor{BrickRed}{\ding{55}} & \textcolor{ForestGreen}{\ding{51}} & \textcolor{BrickRed}{\ding{55}} & \textcolor{ForestGreen}{\ding{51}} & \textcolor{BrickRed}{\ding{55}} \\
{\textbf{Ours}} & {\textbf{A comprehensive binary Q\&A dataset created with PCB design files, spanning 3}\\\textbf{categories of design questions, to enable LLM evaluations}} & \textcolor{ForestGreen}{\ding{51}} & \textcolor{ForestGreen}{\ding{51}} & \textcolor{ForestGreen}{\ding{51}} & \textcolor{BrickRed}{\ding{55}} & \textcolor{BrickRed}{\ding{55}}
\end{tblr}

    }
    \label{tab:related_works}
\end{table*}

\subsection{Motivation}

A key research challenge at the intersection of LLMs and PCBs is the lack of standardised evaluation benchmarks for PCB design and analysis tasks. 
Datasets, including the Stanford Question and Answer Dataset (SQuAD) \cite{rajpurkar_know_2018} and Humanity's Last Exam \cite{phan_humanitys_2025}, exist for natural language processing (NLP) tasks.
Similarly, benchmarks like HumanEval \cite{chen_evaluating_2021} for code generation tasks, or VerilogEval \cite{liu_verilogeval_2023} for EDA provide comprehensive evaluation sets for evaluating LLM performance in niche domains. 
However, no readily available datasets exist for evaluating PCB designs using LLMs.

Of the datasets that already exist for PCBs, image-based datasets are useful for visual anomaly detection on an already manufactured PCB \cite{calabrese_application_2025, zheng_printed_2022, xu_printed_2025}, but there is a scarcity of data corresponding to earlier in the flow, like in the early circuit schematic design stages. 
Furthermore, since they are image-based, they are non-trivial to use with language models in their unprocessed form.
There is therefore a need for a text-based PCB dataset that can be used in the pre-manufacturing design stages to enable catching design flaws earlier in the supply chain.
Curating a binary dataset, such as question-answer pairs like QASPER \cite{dasigi_dataset_2021}, can allow one to deterministically evaluate LLMs on their ability to interpret PCB designs using metrics such as binary classification \cite{sun_ics_2026}.

The following research questions guide our work:
\begin{itemize}[leftmargin=*]
    \item \textbf{RQ1}: Can we produce a reliable PCB question-answer dataset for the systematic evaluation of LLMs in interpreting PCB designs?
    \item \textbf{RQ2}: How do LLMs perform when analysing these circuits using standard formats, i.e. schematics as PDFs, or KiCAD design files?
    \item \textbf{RQ3}: Can an information-dense text-based design format for PCB design files improve the ability of LLMs to correctly interpret designs when tasked with answering the dataset's questions?
\end{itemize}

\subsection{Contributions}

The main contributions of this work are summarised as follows:

\begin{itemize}[leftmargin=*]
    \item We introduce\textbf{ PCB-QA}, the first comprehensive Q\&A dataset consisting of 480 question-answer pairs created from early-stage KiCAD design files. We reference multiple representations of PCB design files -- both native exports from KiCAD and our proposed JSON serialisations of netlists and simulated SPICE netlists -- to generate our dataset questions (Section~\ref{sec:pcb_design_formats}). 
    \item The questions are partitioned into three categories: (a) component connection questions, (b) datasheet inquiries on components, or (c) SPICE simulation-based questions to address multiple aspects of PCB designs while evaluating them in text-based form (Section~\ref{sec:dataset}). 
    \item Leveraging the text-based nature of our dataset, we prompt open-weight and closed-weight LLMs with different text-based PCB design file formats and benchmark their interpretation skills against our dataset (Section~\ref{sec:benchmarking}).
    \item Open-source: Our framework, the KiCAD to JSON scripts which generate JSON representations of schematic netlists and SPICE simulation files, and the PCB-QA dataset are released at \textbf{\url{https://anonymous.4open.science/r/pcb\_qa-4FEE}}.

\end{itemize}

\section{Related Works}

This section reviews publicly available PCB datasets and benchmarking frameworks (categorised in Table~\ref{tab:related_works}). No identified dataset provides textual scenarios covering circuit design and analysis.

\subsection{PCB datasets}

Modern PCB manufacturing pipelines integrate optical and X‑ray inspection tools at each stage to detect surface and assembly‑level defects \cite{ghosh_advanced_2025}. 
Consequently, most PCB datasets for research record \textit{image‑based anomalies} \cite{tang_online_2019, mehta_fics_2022} rather than schematic or netlist‑level design flaws.
These datasets do not specifically
validate adherence to PCB design specifications -- for example, they do not check if a given PCB uses correct port assignments, nor if a design produces functionally consistent circuits before manufacturing. 
These datasets are therefore unsuitable for evaluating LLMs on early design tasks such as schematic interpretation or netlist creation.

\subsection{PCB benchmarks}

While question-answering benchmarks exist in the field of electrical and electronics engineering \cite{meshram_electrovizqa_2024, skelic_circuit_2025, li_eeebench_2025}, they focus on evaluating multimodal LLMs as their prompts incorporate design schematics as images. 
The first benchmark specifically targeting LLM-driven PCB tasks is PCBBench \cite{li_pcbbench_2025}, which evaluates placement and routing optimisation, but this does not assess the preceding circuit design stage, where schematic errors and netlist vulnerabilities originate. 

To date, there is no standardised text-based benchmark which evaluates PCB designs at the netlist or schematic level. Such a contribution would be a first step towards integrating LLMs in the PCB design life cycle.

\begin{figure*}[t]
    \centering
    \includegraphics[width=0.9\linewidth]{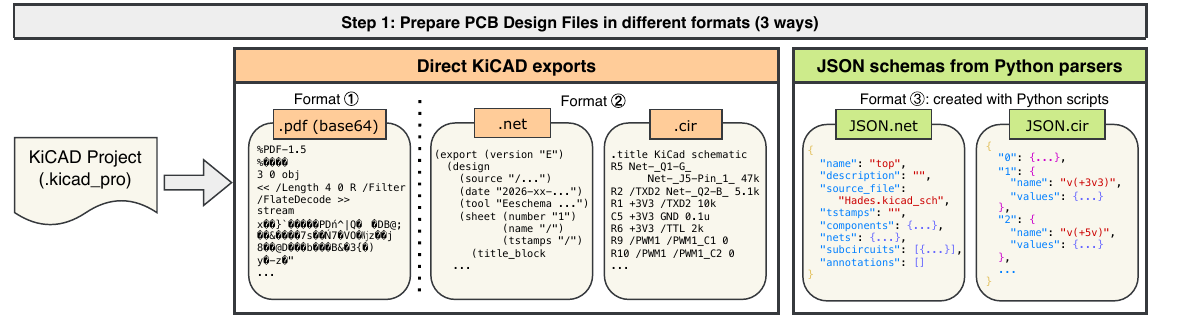}
    \caption{Different format sets of PCB design files. After Step 1 of Figure~\ref{fig:main_figure}, to provide a comprehensive overview of the PCB design, we group them into 2 format sets. Set \ding{192} denotes the base64-encoded schematic file (with binary contents) as a PDF. Set \ding{193} is a collection of the native KiCAD exported files. Set \ding{194} consists of schemas generated by Python scripts on Set \ding{193} to generate simplified JSON-based representations.}
    \label{fig:different_input_formats}
\end{figure*}

\section{PCB Design Formats} \label{sec:pcb_design_formats}

PCB files can be represented in numerous different formats, many of them visual. %
Figure~\ref{fig:different_input_formats} describes the contents of each of the format sets we selected for evaluation in this work. We base our analysis around the open-source PCB editor KiCAD.

\subsection{KiCAD-exported schematic as PDF}

Schematic design files are often interchanged in visual formats, such as PDFs, primarily because such formats are not restricted to a specific CAD editor.
As modern LLMs are capable of ingesting PDF files, it follows that PDF-exported schematics can be considered, perhaps na\"ively, as a suitable machine representation.
We therefore begin our study by exporting PDFs from KiCAD and providing them to an LLM's API in base64 format. In Figure~\ref{fig:different_input_formats}, we denote this file format as design format set \ding{192}.

\subsection{KiCAD-exported netlist files}

\textbf{Schematic netlists}: Netlists exported from KiCAD contain all the relevant metadata of the design. 
However, schematic netlists do not denote the behaviour of components. 
On its own, information like calculating the voltage drop across a net cannot be determined from schematic netlists.

\textbf{SPICE netlists}: SPICE netlists, however, outline the connections of a PCB in text-based formats.
While these files cannot always be directly simulated using `ngspice', components that require specific tunable parameters can be manually defined as \textit{models} within `ngspice' to run simulations and analyse the analog behaviour of PCB designs. 
When used in conjunction with the schematic netlist, these files provide insights into the core design and behaviour of a PCB.
This combination of files forms design format set \ding{193}. %

\subsection{KiCAD design files to a JSON-based format}

\textbf{Schematic netlists:}
KiCAD schematic files for complex PCB designs span across a series of pages to maintain modularity.
Each sheet contains a few top-level components grouped together to perform a particular function.
Associated nets usually cross over between sheets depending on their functionality. 
Therefore, we consider each page as its own subcircuit and the top-level sheet as the top-level circuit.

To represent the main top-level circuit, we require a structured text-based format that can nest other components or subcircuits within it. 
Since LLMs are familiar with interpreting specific instructions from JSON formats, and outputting responses in JSON-schema \cite{madaan_language_2022}, we create a Python parser to convert the design files to a JSON-based format, preserving hierarchical component and net relations.
Our JSON representations for schematic netlists follow the schema outlined in an open-source Pythonic library called CircuitSynth \cite{mattner_circuitsynth_2025}. Our hypothesis is that this representation, which abstracts unnecessary details (e.g. graphical component placements in the KiCAD files, symbol representations, etc.) will provide fewer `distractions' for the LLM when considering the information needed to answer questions correctly.

\textbf{SPICE netlists:}
The .cir file exported from KiCAD is an input to the ngspice simulation engine, but does not itself contain the results of that simulation. For an LLM to make use of this file, it must know how to make use of the engine (e.g. via a plugin or tool call). 
To simplify this task, our conversion script converts the analog and passive components of the netlist into `ngspice'-compliant formats, ignoring any integrated circuits or controller-based modules. 
We then perform this `reduced' simulation automatically, and provide the simulation results in a JSON-formatted output ready for use in LLM prompting. %
This script also appends a transient response block to the circuit to record the voltage and current simulations of each circuit. 
Our hypothesis is that these computed simulation values will help LLMs establish ground truth for questions pertaining to analog component behaviours.

\section{PCB-QA: Dataset Creation Rules} \label{sec:dataset}

To generate our dataset, we have obtained KiCad projects
from different open-source hardware projects.
Table~\ref{tab:no_of_components_nets_oshp} outlines the applications of the selected projects and records their complexities based on how many components and nets they contain. 
Care has been taken to obtain an array of projects from various domains and with differing complexities.  

\begin{figure}[h]
    \subfigure[By question category]
    {
        \includegraphics[width=0.30\textwidth]{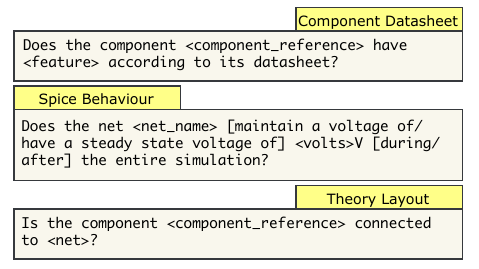}
        \label{fig:question_templates}
    }
    \hspace{-1em}
    \subfigure[For LLM response]
    {
        \includegraphics[width=0.16\textwidth]{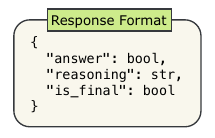}
        \label{fig:llm_response_format}
    }
    \caption{Templates for (a) each question category in the dataset, and (b) the expected response format for LLMs}
    \label{fig:sample_subfigures}
\end{figure}

For each project, all the KiCAD files (including SPICE simulations of the circuits) can be referenced to craft the design question-answer pairs.
Following Step 2 denoted in Figure~\ref{fig:main_figure}, we broadly categorise these questions into: \ding{202} `theory\_layout', \ding{203} `component\_datasheet', and \ding{204} `spice\_behaviour'.
The question templates for each category are defined in Figure~\ref{fig:question_templates}.

\ding{202} \textbf{theory\_layout:} Any questions about a selected component and its associated pin connections fall under this umbrella of questions. 
An example from the `fan\_controller' project is: ``Is the component U1 connected to GND?" 
The flow to answer this question manually is by examining the KiCAD schematic for the `fan\_controller' project or the JSON representation of the `fan\_controller' circuit. 

Going the JSON route, for the net in question (here, it is GND), we determine whether it is in the top-level circuit, or in one of the subcircuit JSONs.
If the net is not found in any circuit or subcircuit, the answer is by default \textit{False}. 
In this case, the GND net is in the top-level circuit.
The list of connections for GND has U1 has one of its entries. 
Therefore, the answer to this question is \textit{True}.

\begin{table}[t]
    \centering
    \caption{Breakdown of the selected open source hardware projects and their applications, along with their component and net numbers.}
    \label{tab:no_of_components_nets_oshp}
    \resizebox{0.45\textwidth}{!}{\definecolor{Concrete}{rgb}{0.949,0.949,0.949}
\begin{tblr}{
  row{even} = {Concrete},
  cell{1}{1} = {r=2}{},
  cell{1}{2} = {r=2}{},
  cell{1}{3} = {c=2}{c},
  vline{2,3} = {-}{0.12em},
  vline{4} = {2-10}{},
  hline{1,3,11} = {-}{0.12em},
  hline{2} = {3-4}{},
  hline{4-10} = {-}{0.04em},
}
Project Name & Application & Number of & \\
 &  & Components & Nets\\
{Acorn Robot\\Electronics\cite{alexander_twistedfields_2021}} & {Solar-powered farming rover/robot with steering\\and drive motors per wheel} & {262} & {223}\\
{HadesFCS\cite{salmony_pms67_2019}} & {Flight controller system with peripheral support} & {206} & {254} \\
{OPNHydro-r2\cite{vonk_opnhydro_2026}} & {Hydroponic systems controller, monitoring soil\\pH, humidity, air temperature and remote access} & {148} & {85} \\
{CF-Chef\cite{reilley_reilleya_2025}} & {Controller for custom composite curing ovens} & {55} & {48} \\
{Meshinger\cite{awgh_meshinger_2022}} & {Handheld wireless router with battery,\\eInk display, SD card and keyboard} & {35} & {73} \\
{Stack-Chan \cite{ishikawa_stackchan_2021}} & {Interactive robot programmed to emote for\\communicating with humans} & {33} & {40} \\
{Portal Hard-\\-ware Wallet\cite{filini_twentytwohw_2024}} & {Compact NFC-based hardware wallet to secure,\\store and transfer Bitcoin} & {31} & {34}\\
{ATTiny Fan\\Controller\cite{arya_crimier_2019}} & {Simple PWM fan controller for 5V\\or 12V fans with I2C support or\\ potentiometer-based control} & {25} & {15} \\
\end{tblr}
}
\end{table}

\ding{203} \textbf{component\_datasheet:} Questions of this category involve extracting important contextual information from the design question, searching through a datasheet, finding the relevant information and retrieving an answer with that information. 
Simple electronic configuration information (like minimum or maximum operating temperatures, maximum acceptable current, or threshold voltage) fall under this category of questions. 
The related keywords can be examined in the datasheet and determined as \textit{True} or \textit{False}. 
To script this retrieval, datasheet embeddings can be used, which is discussed in Section~\ref{subsec:prompt_design}.

\ding{204} \textbf{spice\_behaviour:} Based on the numerical simulation values generated by `ngspice', questions can be asked about the voltage or current metrics against any net in the circuit. 
An example from the `stack-chan' project is: ``Does the /pwm1 signal maintain a voltage of 0V during the entire simulation?"
Such questions can be answered either by manually simulating the SPICE circuit from KiCAD, by inputting certain parameters, or looking at the parsed SPICE simulations in the JSON dictionary.
In its SPICE JSON, we see 19 simulated variables and 108 data points. 
Since the keywords are `maintaining voltage', this question implies the steady state voltage of the `/pwm1' signal, which can be approximated as the average of the last $\sim$10 data points for `/pwm1'.
In the JSON file, the range falls around $\sim$3.46V (with a precision of 2 decimal places), not 0V. 
Hence, the answer is \textit{False}. 

We note that nets that are not simulated may not appear in the SPICE simulated JSON, and this SPICE JSON file is not exhaustive, but it still offers insights into circuit functionality in ways that schematics alone cannot.

All the 480 design question-answer pairs that form the dataset have been manually created to maintain question integrity and ensure correctness in responses. This produced artifact answers RQ1.
While every project consists of an equal distribution of questions from the 3 categories, the questions crafted do not have an equal split of "YES" and "NO" responses -- this does make the dataset in its current form imbalanced. Future work will balance the questions to have an equal split of each. 

\section{PCB-QA: Benchmarking LLMs} \label{sec:benchmarking}

\subsection{Evaluation metrics}
We have designed our question-answer pairs to only have two possible answers: "YES" (True) or "NO" (False).
We therefore choose standard binary classification metrics to measure model performance against our dataset. 
`Accuracy', `Precision', `Recall' and `F1-Score' \cite{hu_unveiling_2024} metrics can be obtained directly from the \textit{sklearn} module defined in Python. 

For imbalanced datasets, accuracy alone is not a reliable predictor. 
The ability of a model to also \textit{correctly predict} a response (i.e. precision) and to find all instances of those predictions (i.e. recall) are taken into consideration. 
Hence, the F1-score normalises the deviations in precision and recall and provides insights into the reliability of the models for the imbalanced dataset.

\subsection{Prompt design} \label{subsec:prompt_design}

As an initial test of how LLMs interpret different input design formats, we select a state-of-the-art LLM that supports file inputs `GPT-5.4-Nano' as a trial.
To aid in response interpretation, we define the response format as depicted in Figure~\ref{fig:llm_response_format}. 
As they are binary-choice, this method allows us to straightforwardly automate checking of LLM responses to the dataset questions.

We therefore feed our question, our design format set and our expected response format to the `GPT 5.4 Nano' model and observe its responses for the first 25\% (a randomised split) of our dataset.
Table~\ref{tab:gpt-5.4-nano_3_prompt_types} depicts `GPT 5.4 Nano's responses against our established classification metrics, where PDF denotes a PDF-format schematic exported from KiCAD; NNet and NCir denote native KiCAD exports (from format set \ding{193}); and PNet and PCir highlight the proposed JSON file format (format set \ding{194}).

\begin{table}[h]
    \centering
    \caption{Evaluation of a 25\% split of our dataset on a state-of-the-art LLM: GPT 5.4 Nano with 3 different input formats. NNet and NCir denote native KiCAD exports (format set \ding{193}) while PNet and PCir denote the proposed JSON-format files (format set \ding{194}).} 
    \resizebox{0.55\linewidth}{!}{
    \definecolor{Concrete}{rgb}{0.949,0.949,0.949}
\begin{tblr}{
  row{even} = {Concrete},
  row{2} = {c},
  cell{1}{1} = {r=2}{},
  cell{1}{2} = {c=3}{c,font=\bfseries},
  cell{3-6}{2-4} = {r},
  vline{2} = {-}{0.12em},
  vline{3-4} = {2-6}{},
  hline{2} = {-}{},
  hline{1,3,7} = {-}{0.12em}
}
{\textbf{Metrics/Input}\\\textbf{Format}} & GPT 5.4 Nano   &  & \\
 & {PDF\\\ding{192}} & {NNet \&\\NCir \ding{193}} & {PNet \&\\PCir \ding{194}}\\
Accuracy & 0.43 & 0.55 & \textbf{0.88}\\
Precision & 0.93 & 0.98 & 0.97\\
Recall & 0.17 & 0.29 & 0.84\\
F1 Score & 0.28 & 0.45 & \textbf{0.90}
\end{tblr}

    }
    \label{tab:gpt-5.4-nano_3_prompt_types}
\end{table}

\textbf{Initial takeaways:} \label{subsubsec:llm_tool_set}
Firstly, we note that the number of input tokens required for this initial experiment rises to the 100k mark when using format sets \ding{193} and \ding{194}, especially for AcornRobotElectronics \cite{alexander_twistedfields_2021} and HadesFCS \cite{salmony_pms67_2019} which are the most complex projects in our dataset according to Table~\ref{tab:no_of_components_nets_oshp}. This motivates us to find ways to compress inputs: instead of flooding context windows with design information that may be irrelevant to a question, we can filter data based on hints from the question.
Python helper functions are defined as tools that the LLM can use to filter for relevant component and net subcircuits from the JSON formats, and relevant schematic netlist and SPICE circuit snippets from the KiCAD exports. 

Secondly, we note that the LLM can answer questions over each format correctly (Ans. RQ2), but the expensive PDF-input type performs much worse than the text-based file inputs. As such, for the rest of our experiments, we disregard PDF-inputs. This has the additional benefit of allowing more diverse LLM experimentation (not all LLMs support encoded file inputs.)
The PDF-specific performance is discussed further in Section~\ref{subsec:prompt_with_pdfs}. 

\begin{table*}[b]
    \centering
    \caption{Binary classification metrics of the selected LLMs on a 25\% subset of the entire dataset, with different input combinations from design sets \ding{193} and \ding{194}. NNet and NCir denote native KiCAD exports, PNet and PCir the proposed JSON-format files.}
    \resizebox{0.98\linewidth}{!}{
    \definecolor{Concrete}{rgb}{0.949,0.949,0.949}
\begin{tblr}{
  row{even} = {Concrete},
  row{1} = {c,font=\bfseries},
  cell{1}{1} = {r=2}{},
  cell{1}{2} = {c=4}{},
  cell{1}{6} = {c=4}{},
  cell{1}{10} = {c=4}{},
  cell{1}{14} = {c=4}{},
  cell{3-6}{2-17} = {r},
  hline{1,3,7} = {-}{0.12em},
  hline{2} = {-}{},
  vline{2,6,10,14} = {-}{}
}
Metrics/Models & {Claude Sonnet 4.6} &  &  &  & {Gemini 3 Flash Preview} &  &  &  & {GPT 5.4 Nano} &  &  &  & {Llama 3.3 70B Instruct} &  &  & \\
 & {NNet \&\\NCir \ding{193}} & {NNet \&\\PCir} & {PNet \&\\NCir} & {PNet \&\\PCir \ding{194}} & {NNet \&\\NCir \ding{193}} & {NNet \&\\PCir} & {PNet \&\\NCir} & {PNet \&\\PCir \ding{194}} & {NNet \&\\NCir \ding{193}} & {NNet \&\\PCir} & {PNet \&\\NCir} & {PNet \&\\PCir \ding{194}} & {NNet \&\\NCir \ding{193}} & {NNet \&\\PCir} & {PNet \&\\NCir} & {PNet \&\\PCir \ding{194}}\\
Accuracy & 0.75 & 0.79 & 0.69 & 0.91 & 0.83 & 0.83 & 0.70 & \textbf{0.92} & 0.48 & 0.60 & 0.68 & 0.89 & 0.45 & 0.62 & 0.53 & 0.79\\
Precision & 0.98 & 0.98 & 0.98 & 0.99 & 0.97 & 0.98 & 0.98 & 0.99 & 1 & 0.95 & 0.98 & 0.99 & 0.96 & 0.8 & 0.96 & 0.98\\
Recall & 0.65 & 0.71 & 0.56 & 0.88 & 0.77 & 0.76 & 0.58 & 0.89 & 0.24 & 0.44 & 0.55 & 0.85 & 0.33 & 0.58 & 0.33 & 0.71\\
F1 Score & 0.78 & 0.82 & 0.71 & 0.93 & 0.86 & 0.86 & 0.73 & \textbf{0.94} & 0.39 & 0.60 & 0.70 & 0.92 & 0.49 & 0.73 & 0.49 & 0.82
\end{tblr}

    }
    \label{tab:25percent_dataset_split}
\end{table*}

\textbf{Datasheet embeddings}:
We ensure that for every project -- if components are referenced in the 'component\_datasheet' questions -- we download the specific datasheets and save them locally. 
For instance: if a question for component `C4' needs a datasheet lookup, a PDF titled `C4.pdf' will already exist in the `datasheets/' directory prior to the LLM run.

On the downloaded datasheets, we generate embeddings by chunking its contents and storing the obtained vectors. 
The \textit{`all-MiniLM-L6-v2'} model \cite{huggingface_sentencetransformers_2024} from the SentenceTransformer module in Python is used to create the vectors and perform semantic searches.
`component\_datasheet' questions can be manually answered by mapping the keywords in the question to those of the datasheet. 
We expect the required contextual information (to support our answer) to be contained in one sentence, so a chunk size of 500 characters and a chunk overlap of 50 characters is sufficient for our runs.
The LLM can then generate an answer to the inputted question with this baseline knowledge included as input. 

\subsection{Selected LLMs}

The LLMs we choose must primarily satisfy two constraints. 
Firstly, the models we select must support integration of our defined tools from Section~\ref{subsubsec:llm_tool_set}.
We want 
the LLM to decide which tool to use for its response generation. 
Secondly, the models must offer reasoning 
to confirm they are not hallucinating their answers.

Keeping these criteria in mind, we have selected: `Claude Sonnet 4.6' \cite{anthropic_claude_2026}, which scores 89.9\% on the GPQA Diamond benchmark, `Gemini 3 Flash Preview', which has a 1M token context window \cite{google_gemini_2025}, and the open-source `Llama 3.3 70B Instruct' model for its lowest cost reasoning at \$0.15/1M tokens \cite{meta_metallama_2024} while still achieving 88.4\% on the HumanEval benchmark. 
In addition, we also pick `GPT 5.4 Nano' for its tool integration support, latency of 2.68 seconds \cite{bristot_compare_2025} and since we attempted our proof-of-concept run with this model.

\section{Results and Discussions} \label{sec:results}

In this section, we record the results we obtain by prompting our selected LLMs with the design format sets \ding{192}, \ding{193} and \ding{194} and observe deviations in these results depending on the category of questions the LLMs evaluate.

\begin{table*}[t]
    \centering
    \caption{Binary classification metrics of the selected LLMs prompted with format sets \ding{193} and \ding{194} on the entire dataset. NNet and NCir denote native KiCAD exports (format set \ding{193}) while PNet and PCir denote our proposed JSON-format files (format set \ding{194}).}
    \resizebox{0.6\linewidth}{!}{\definecolor{Concrete}{rgb}{0.949,0.949,0.949}
\begin{tblr}{
  row{even} = {Concrete},
  row{1} = {c,font=\bfseries},
  cell{1}{1} = {r=2}{},
  cell{1}{2} = {c=2}{},
  cell{1}{4} = {c=2}{},
  cell{1}{6} = {c=2}{},
  cell{1}{8} = {c=2}{},
  cell{3-6}{2-10} = {r},
  cell{2}{2-9} = {r},
  hline{1,3,7} = {-}{0.12em},
  hline{2} = {-}{},
  vline{2,4,6,8} = {-}{}
}
Metrics/Models & {Claude Sonnet 4.6} &  & {Gemini 3 Flash Preview} &  & {GPT 5.4 Nano} &  & {Llama 3.3 70B Instruct} & \\
 & {NNet \&\\NCir \ding{193}} & {PNet \&\\PCir \ding{194}} & {NNet \&\\NCir \ding{193}} & {PNet \&\\PCir \ding{194}} & {NNet \&\\NCir \ding{193}} & {PNet \&\\PCir \ding{194}} & {NNet \&\\NCir \ding{193}} & {PNet \&\\PCir \ding{194}}\\
Accuracy & 0.79 & 0.90 & 0.86 & \textbf{0.93} & 0.55 & 0.88 & 0.48 & 0.84\\
Precision & 0.96 & 0.97 & 0.97 & 0.97 & 0.98 & 0.97 & 0.96 & 0.97\\
Recall & 0.70 & 0.87 & 0.80 & 0.92 & 0.29 & 0.84 & 0.33 & 0.78\\
F1 Score & 0.80 & 0.92 & 0.88 & \textbf{0.95} & 0.45 & 0.90 & 0.49 & 0.86
\end{tblr}
}
    \label{tab:100percent_dataset}
\end{table*}

\subsection{GPT 5.4 Nano on PDFs (format set \ding{192})} \label{subsec:prompt_with_pdfs}
(GPT 5.4 Nano)'s performance on the first 25\% of our dataset is presented in Table~\ref{tab:gpt-5.4-nano_3_prompt_types}.
From the 3 format sets, format set \ding{192} (PDFs) has the lowest accuracy and F1 scores for the split. 
We analyse this further with Table~\ref{tab:gpt-5.4-nano_category_wise}, which outlines the category-wise evaluation of format set \ding{192} on the same subsection (first 25\%) of the dataset.

\begin{table}[h]
    \centering
    \caption{Sub-evaluation of format set \ding{192} in the 25\% dataset split on GPT 5.4 Nano against our 3 question categories.}
    \resizebox{0.7\linewidth}{!}{
    \definecolor{Concrete}{rgb}{0.949,0.949,0.949}
\begin{tblr}{
  row{even} = {Concrete},
  row{2} = {c},
  cell{1}{1} = {r=2}{},
  cell{1}{2} = {c=3}{c,font=\bfseries},
  cell{3-6}{2-4} = {r},
  vline{2} = {-}{0.12em},
  vline{3-4} = {2-6}{},
  hline{2} = {-}{},
  hline{1,3,7} = {-}{0.12em}
}
{\textbf{Metrics/Category}} & GPT 5.4 Nano on \ding{192}  &  & \\
 & {component\_\\datasheet} & {spice\_\\behaviour} & {theory\_\\layout}\\
Accuracy & 0.40 & 0.33 & 0.57 \\
Precision & 1 & 0 & 0.93 \\
Recall & 0.04 & 0 & 0.46 \\
F1 Score & 0.07 & 0 & 0.62 \\
\end{tblr}

    }
    \label{tab:gpt-5.4-nano_category_wise}
    \vspace{-5mm}
\end{table}

We find that `theory\_layout' questions are answered with the highest accuracy of 57\% and F1 score of 62\%, while `spice\_behaviour' questions are answered with 33\% accuracy but an F1 score of 0\%.
This is because while the schematic denotes what components and nets are connected together, enabling insights for `theory\_layout' questions, schematics do not store or depict any information on the SPICE behaviour of the design.
What is interesting, however, is that `component\_datasheet' queries are answered with 40\% accuracy and a precision of 100\%. 
This means GPT 5.4 Nano is able to obtain component information from the PDF, search for the datasheet either using the web or its pretrained knowledge, and look through it for the answer. 
However, the F1 score for this category is only 7\%, indicating the model outputs a high number of false negatives.
These results prove that (a) format set \ding{192} is not comprehensive and does not cover other aspects of PCB design, and  that (b) accuracy alone is insufficient to measure LLM performance for our dataset.

\subsection{Ablation on format sets \ding{193} and \ding{194}}

For cost, we use the first 25\% of the dataset to measure how the sub-components of format sets \ding{193} and \ding{194} can impact model performance.
Here, the LLMs should call the appropriate tools and provide the correct final answer.
These ablations are recorded in Table~\ref{tab:25percent_dataset_split}. 
We see that for every model, format set \ding{194} has the highest accuracy and F1 scores, indicating that our proposed JSON formats (1) can offer design context in a format LLMs can interpret and (2) these files can be used to reliably answer questions pertaining to several aspects of PCB design. 

\subsection{Full dataset evaluation by selected LLMs}

Having validated our proposed JSON format set against the 25\% subset, we now present the more expensive run on the entire set of 480 questions spanning the 3 question categories, comparing format set \ding{193} and \ding{194}. 
We tabulate our findings in Table~\ref{tab:100percent_dataset}, which show Gemini 3 Flash Preview outperforms the other evaluated LLMs, both for the native KiCAD format set \ding{193} and the proposed JSON format \ding{194}, scoring an accuracy of 93\% for the latter while answering RQ2 and RQ3. 

\begin{figure}[h]
    \centering
    \resizebox{0.7\linewidth}{!}{
    \begin{tikzpicture}
 \pgfplotsset{%
    width=\linewidth,
    height=0.9\linewidth    
}
\begin{axis}[
    ybar,
    ymin=0.5, 
    ymax=1.2,
    bar width=9pt,
    xlabel={Category},
    ylabel={Accuracy},
    symbolic x coords={CD, SB, TL},
    xtick=data,
    visualization depends on={x \as \xcoord},
    nodes near coords,
    every node near coord/.append style={
        font=\footnotesize,
        rotate=45,
        color=black,
        inner sep=5pt,
        xshift=5pt
    },
    legend image code/.code={
        \draw[#1, fill=#1] 
        (0cm,-0.1cm) rectangle (0.2cm,0.1cm);
    },
    legend style={
        at={(0.25,0.7)}, 
        anchor=south, 
        font=\footnotesize, 
        nodes={inner xsep=3pt}, 
        legend cell align=left, 
        draw=none, 
        scale=0.5, 
        transform shape, 
        legend columns=1
    },
    enlarge x limits=0.25,
]

\addplot[fill=Melon!60, point meta=explicit symbolic] coordinates {
    (CD, 0.7875) [0.79]
    (SB, 0.94375)
    (TL, 0.975) 
};

\addplot[fill=SpringGreen!75] coordinates {
    (CD, 0.88125) 
    (SB, 0.94375) 
    (TL, 0.975)
};

\addplot[fill=Yellow!50, point meta=explicit symbolic, every node near coord/.append style={xshift=6pt, inner sep=1pt}] coordinates {
    (CD, 0.71875) [0.72]
    (SB, 0.94375) 
    (TL, 0.975) 
};

\addplot[fill=Gray!30, point meta=explicit symbolic, every node near coord/.append style={xshift=6pt, inner sep=1pt}] coordinates {
    (CD, 0.65625) [0.66]
    (SB, 0.9) [0.90]
    (TL, 0.975) 
};

\legend{claude-sonnet-4.6, gemini-3-flash-preview, gpt-5.4-nano, llama-3.3-70b-instruct}

\end{axis}
 
\end{tikzpicture}}
    \vspace{-0.5em}
    \caption{Accuracy of responses per model, evaluated against our 3 question categories on format set \ding{194}. CD stands for `component\_datasheet', SB for `spice\_behaviour' and TL for `theory\_layout'.}
    \label{fig:category_wise_accuracy}
    \vspace{-5mm}
\end{figure}
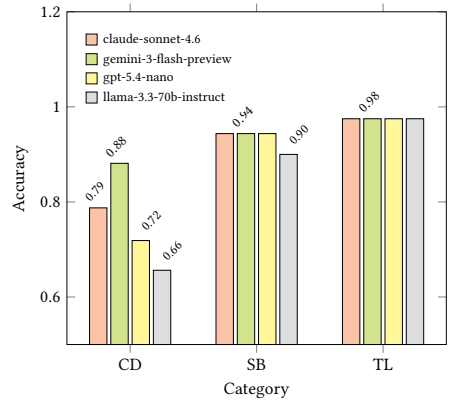

\textbf{Category-wise comparisons of the selected LLMs}:
Finally, we compare the performances of the selected LLMs against our proposed format set \ding{194}.
Figure~\ref{fig:category_wise_accuracy} depicts the accuracy of each LLM on each of the 3 categories of questions (denoted by CD for `component\_datasheet', SB for `spice\_behaviour' and TL for `theory\_layout'). 

Using format set \ding{194}, the number of accurate LLM responses is highest for the `theory\_layout' category, where connections can be determined by querying the PNet. 
`spice\_behaviour' answers are also highly accurate, as the LLMs can query PCir for the SPICE simulation values. 
However, for the `component\_datasheet' category, we record notably lower values of correct responses, varying drastically from model to model as `component\_datasheet' questions are the only type of questions that require semantic search to retrieve context. 

\textbf{Method limitations}: Our prompting method for the native KiCAD format set \ding{193} does not optimise for input token size, and we do not perform checks to ensure the input we provide fits into the LLM context window. 
Given this, we note that the Llama 3.3 70B Instruct model rejected API calls for the `AcornRobotElectronics' project, which is the most complex project in our dataset according to Table~\ref{tab:no_of_components_nets_oshp}. %
The error it outputs is rendered invalid, as it does not match the response format we provide (Figure~\ref{fig:llm_response_format}) and is hence excluded from aggregation.

\section{Conclusion} \label{sec:conclusion}
This paper presents the first open-source question-answer dataset for PCB designs consisting of manually created 480 design questions obtained from 8 different open-source hardware projects with varying complexities.
The dataset contains questions spanning 3 different categories to analyse PCB designs on multiple aspects.
Using this dataset, we evaluated selected state-of-the-art LLMs across different PCB file format representations to analyse their interpretation capabilities. 
We observe that from the selected LLMs, Gemini 3 Flash Preview has the highest accuracy of 93\% when prompted with the proposed JSON-based format, demonstrating pathways for reliable LLM interpretation in this setting.
To the best of our knowledge, this is the first text-based comprehensive PCB dataset that evaluates LLM at the netlist and schematic design stage.

\bibliographystyle{ACM-Reference-Format}
\bibliography{refs/29_may_MLCAD}

\end{document}